\documentclass[12pt]{article}
\usepackage{a4wide,epsfig,amsmath,amssymb,cite,slashed}
\usepackage{scalefnt}
\begin{document}

%%%%%%%%%%%%%%%%%%%%%%%%%%%%%%%%%%%%%%%%%%%%%%%%%%%%%%%%%%%%

\title{\vskip-3cm{\baselineskip14pt
\begin{flushleft}
\normalsize DESY 06-031 \\
\normalsize SFB/CPP-06-12 \\
\normalsize TTP06-11 \\
\end{flushleft}}
\vskip1.5cm
Two-Loop Matching Coefficients for Heavy Quark Currents}

\author{\small B.A. Kniehl$^{(a)}$, A. Onishchenko$^{(a,b)}$,
  J.H. Piclum$^{(a,c)}$, M. Steinhauser$^{(c)}$
\\
{\small\it (a) II. Institut f\"ur Theoretische Physik, 
  Universit\"at Hamburg}\\
{\small\it 22761 Hamburg, Germany}
\\
{\small\it (b) Theoretical Physics Department,
  Petersburg Nuclear Physics Institute, Orlova Roscha}\\
{\small\it 188300 Gatchina, Russia}
\\
{\small\it (c) Institut f{\"u}r Theoretische Teilchenphysik,
  Universit{\"a}t Karlsruhe}\\
{\small\it 76128 Karlsruhe, Germany}
}

\date{}

\maketitle

\thispagestyle{empty}

\begin{abstract}
In this paper we consider the matching coefficients up to two loops 
between Quantum Chromodynamics (QCD) and Non-Relativistic QCD (NRQCD)
for the vector, axial-vector, scalar and pseudo-scalar currents.
The structure of the effective theory is discussed and analytical
results are presented. Particular emphasis is put on the singlet
diagrams.

\medskip

\noindent
PACS numbers:
\end{abstract}

%\newpage

%- }}}
%- {{{ main text:

In the recent years quite a lot of activity has been devoted to the
treatment of bound states of two heavy particles both in QED and 
QCD (for a recent review see, e.g., Ref.~\cite{Brambilla:2004wf}).
From the theory point of view the calculations have been put onto a
solid basis due to the formulation of proper effective
theories~\cite{Caswell:1985ui,Bodwin:1994jh}, NRQED and NRQCD,
respectively, which provide the 
possibility to systematically evaluate higher order corrections.
The construction of the effective theories consists of essentially two
steps: First, the effective operators involving the light degrees of
freedom have to be constructed and second the corresponding couplings,
the so-called coefficient functions, have to be computed by comparing
the full and the effective theories. The latter is also referred to as
matching calculation.

The framework which is considered in this letter consists of QCD
accompanied by external currents where we allow for 
vector, axial-vector, scalar and pseudo-scalar couplings.
The main results of this letter are the two-loop matching
coefficients.
Thus, following the prescription outlined above we determine in a
first step the effective currents and then perform a matching
calculation.

The matching coefficients provided in this paper constitute a building
block in all calculations involving the corresponding external
currents. This includes in particular 
production and decay processes of heavy quarkonia or the production of
top quark pairs close to threshold.
One could also think of the decay of a CP-even or CP-odd Higgs boson
(with mass $M$) into two quarks with $2 m\approx M$.

The basic idea behind the construction of the 
Lagrange density for NRQCD is to expand all terms 
of the QCD Lagrangian in the limit of a large quark mass.
A similar procedure has to be applied to external currents which we
define in coordinate space as
\begin{eqnarray}
  j_v^\mu &=& \bar{\psi} \gamma^\mu \psi\,, \nonumber \\
  j_a^\mu &=& \bar{\psi} \gamma^\mu\gamma_5 \psi\,, \nonumber \\
  j_s     &=& \bar{\psi} \psi\,, \nonumber \\
  j_p     &=& \bar{\psi} i\gamma_5 \psi\,.
  \label{eq::currents}
\end{eqnarray}
Note that the anomalous dimension of $j_v^\mu$ and $j_a^\mu$ is zero
whereas for the scalar and pseudo-scalar current it is obtained from
the renormalization constant $Z_s = Z_p = Z_m$, which is given
at the two-loop level in Ref. \cite{Gray:1990yh}.

In order to perform the transition to the effective theory it is
convenient to work in momentum space and 
to introduce the two-component Pauli-spinors
in the form
\begin{equation}
  u(\vec{p}\,) = \sqrt{\frac{E+m}{2 E}} \left(
  \begin{array}{c}
    \chi \\ \frac{\vec{p}\cdot\vec{\sigma}}{E+m} \chi
  \end{array}
  \right)\,,\;
  v(-\vec{p}\,) = \sqrt{\frac{E+m}{2 E}} \left(
  \begin{array}{c}
    \frac{(-\vec{p}\,)\cdot\vec{\sigma}}{E+m} \phi \\ \phi
  \end{array}
  \right)\,,
  \label{eq::psi2phi}
\end{equation}
where $m$ denotes the heavy quark mass. In Eq.~(\ref{eq::psi2phi})
$\chi$ is a spinor that annihilates a heavy quark and 
$\phi$ correspondingly creates a heavy anti-quark with momentum $\vec{p}$.

In a first step we want to express the currents of Eq.~(\ref{eq::currents})
in terms of $\phi$ and $\chi$ and expand in the inverse heavy quark
mass. This actually leads to the tree-level matching conditions.
When inserting Eq.~(\ref{eq::psi2phi}) into (\ref{eq::currents})
it turns out to be convenient to split the time-like and
space-like coefficients of the vector and axial-vector currents. This
leads to
\begin{eqnarray}
  j_v^0 &=& 0 + {\cal O}\left(\frac{1}{m^2}\right)\,,\nonumber\\
  j_v^k &=& \tilde{j}_v^k + {\cal O}\left(\frac{1}{m^2}\right)\,,\nonumber\\
  j_a^0 &=& i\tilde{j}_p + {\cal O}\left(\frac{1}{m^2}\right)\,,\nonumber\\
  j_a^k &=& \tilde{j}_a^k + {\cal O}\left(\frac{1}{m^3}\right)\,,\nonumber\\
  j_s   &=& \tilde{j}_s + {\cal O}\left(\frac{1}{m^3}\right)\,,\nonumber\\
  j_p   &=& \tilde{j}_p + {\cal O}\left(\frac{1}{m^2}\right)\,,
\end{eqnarray}
where $k=1,2,3$ and the currents in the effective theory are given by
\begin{eqnarray}
  \tilde{j}_v^k &=& \phi^\dagger \sigma^k \chi \,,\nonumber\\
  \tilde{j}_a^k &=& \frac{1}{2 m} \phi^\dagger
  [\sigma^k,\vec{p}\cdot\vec{\sigma}] \chi \,,\nonumber\\
  \tilde{j}_s &=& -\frac{1}{m} \phi^\dagger \vec{p}\cdot\vec{\sigma}
  \chi \,,\nonumber\\
  \tilde{j}_p &=& -i\phi^\dagger \chi \,.
\end{eqnarray}
Note that $\tilde{j}_p$ also appears in the expansion of $j_a^0$ which
means that the corresponding matching coefficients are equal.
This will be used as a check of our calculation.
Due to the occurrence of the momentum $\vec{p}$ in 
$\tilde{j}_a^k$ and $\tilde{j}_s$ an
expansion in the external momenta has to be performed in order to
obtain the loop corrections to the corresponding matching
coefficients.

The basic idea to obtain the matching coefficients is to compute 
vertex corrections induced by the considered current both in the full
and the effective theory. In practice it is convenient to consider
the renormalized vertex function with two external on-shell quarks and
to perform an asymptotic expansion about $s=4m^2$, where $s$ is the
momentum squared of the external current, the so-called threshold
expansion~\cite{Beneke:1997zp,Smirnov:2002pj}. 
Denoting by $\Gamma_x$ the proper structure of the 
genuine vertex corrections and by
$Z_2$ and $Z_x$ the renormalization constants due to the
quark wave function and the anomalous dimension of the current
one obtains the equation
\begin{eqnarray}
  Z_2 Z_x \Gamma_x(q_1,q_2) &=& c_x \tilde{Z}_2 \tilde{Z}_x^{-1}
  \tilde{\Gamma}_x + \ldots
  \label{eq::match_def}
  \,,
\end{eqnarray}
where $x\in\{v,a,s,p\}$ with the understanding that the 
axial-vector part is split into time-like and space-like
components.
The ellipses denote terms suppressed by inverse powers of the
heavy quark mass and the quantities in the effective theory are marked
by a tilde. $c_x$ is the matching coefficient we are after.
In our approximation $\tilde{Z}_2=1$.  $Z_2$ to two loops has been
computed in Ref.~\cite{Broadhurst:1991fy}. 
As far as $\tilde{\Gamma}_x$ is concerned only the tree-level result
determined by 
$\tilde{j}_x$ contributes to Eq.~(\ref{eq::match_def}).
The momenta $q_1$ and $q_2$ in Eq.~(\ref{eq::match_def}) correspond to
the outgoing momenta of the quark and anti-quark which are considered
on-shell.

Starting from order $\alpha_s^2$ the matching coefficients $c_x$
exhibit infra-red divergences which are compensated by ultra-violet
divergences of the effective theory rendering physical quantities
finite. In Eq.~(\ref{eq::match_def}) the renormalization constant
$\tilde{Z}_x$ which generates the anomalous dimension of $\tilde{j}_x$
takes over this part.

The quantities $\Gamma_x$ are conveniently obtained with the help of
projectors which are constructed in such a way that they project on the
coefficients of $\tilde{\Gamma}_x$.
For the vector case, the zeroth component of the axial-vector 
and the pseudo-scalar case we can simply identify
$q_1^2=q_2^2=q^2/4=m^2$ and use
\begin{eqnarray}
  \Gamma_v &=& \mbox{Tr}\left[ P^{(v)}_{\mu} \Gamma^{(v),\mu} \right]\,,
  \nonumber\\
  \Gamma_p &=& \mbox{Tr}\left[ P^{(p)} \Gamma^{(p)} \right]\,,
  \nonumber\\
  \Gamma_{a,0} &=& \mbox{Tr}\left[ P^{(a,0)}_{\mu} \Gamma^{(a),\mu} \right]\,,
  \label{eq::proj0}
\end{eqnarray}
with
\begin{eqnarray}
  P^{(v)}_{\mu} &=& \frac{1}{8 (D-1) m^2} \left( -\frac{\slashed{q}}{2} + m
  \right) \gamma_\mu \left( \frac{\slashed{q}}{2} + m
  \right)\,,\nonumber\\
  P^{(p)} &=& \frac{1}{8 m^2} \left( -\frac{\slashed{q}}{2} + m
  \right) \gamma_5 \left( \frac{\slashed{q}}{2} + m
  \right)\,,\nonumber\\
  P^{(a,0)}_{\mu} &=& -\frac{1}{8 m^2} \left( -\frac{\slashed{q}}{2} + m
  \right) \gamma_\mu \gamma_5 \left( \frac{\slashed{q}}{2} + m
  \right)\,.
  \label{eq::proj1}
\end{eqnarray}
As already mentioned above  the case $(a,0)$ is used as a check for the
pseudo-scalar matching coefficient.

For the axial-vector and scalar cases we have the equations analogous
to Eq.~(\ref{eq::proj0}). However, since the corresponding 
effective currents have a suppression factor $|\vec{p}\,|/m$ it is
necessary to choose $q_1=q/2+p$ and $q_2=q/2-p$, to expand up to linear
order in $p$ and to set afterwards $p=0$ and $q^2=4m^2$. Note that we
choose a reference frame where $q\cdot p =
0$~\cite{Beneke:1997zp,Smirnov:2002pj}.
Thus the projectors are more complicated and are given by
\begin{eqnarray}
  P_{(a,i),\mu} &=& -\frac{1}{8 m^2} \left\{ \frac{1}{D-1}
    \left(-\frac{\slashed{q}}{2} + m \right) \gamma_\mu \gamma_5
    \left(-\frac{\slashed{q}}{2} + m\right) \right. \nonumber\\
  && \left. - \frac{1}{D-2} \left(-\frac{\slashed{q}}{2} + m\right)
    \frac{m}{p^2} \left( (D-3) p_\mu + \gamma_\mu \slashed{p} \right)
    \gamma_5 \left(\frac{\slashed{q}}{2} + m\right) \right\}
    \,,\nonumber\\
  P_{(s)} &=& \frac{1}{8 m^2} \left\{ \left(-\frac{\slashed{q}}{2} + m
    \right) {\bf 1} \left(-\frac{\slashed{q}}{2} + m\right) +
    \left(-\frac{\slashed{q}}{2} + m\right) \frac{m}{p^2} \slashed{p}
    \left(\frac{\slashed{q}}{2} + m\right) \right\}
    \,.
  \label{eq::proj2}
\end{eqnarray}

\begin{figure}
  \leavevmode
  \epsfxsize=\textwidth
 \epsffile{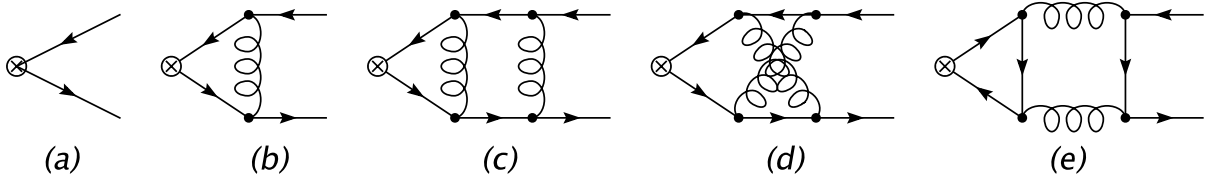}
\caption{\label{fig::diags} Feynman diagrams contributing to the 
  matching coefficients. In (e) the so-called singlet diagram is
  shown which does not contribute to $c_v$. In the closed fermion loop
  all quark flavours have to be considered.}
\end{figure}

In Fig.~\ref{fig::diags} some Feynman diagrams contributing to the
matching coefficients are shown. Due to the application of the
projectors the corresponding integrals can be reduced to the functions
$J_\pm$ and $L_\pm$ as defined in Eqs.~(14) and~(55)
of Ref.~\cite{Beneke:1997zp}. However, due to the expansion in the
momentum $p$ the powers of the denominators are higher and a
systematic reduction of the scalar integrals to master integrals is
necessary. For the current calculation we implemented the method of
Ref.~\cite{Smirnov:2003kc}. For some of the occuring integrals 
the program {\tt AIR}~\cite{Anastasiou:2004vj} is applied. 
The details will be described elsewhere.

An important class of diagrams is constituted by the
so-called singlet diagrams (cf. Fig.~\ref{fig::diags}(e)) where the
external current does not couple to the quark--anti-quark pair of the
final state. Due to Furry's theorem there is no contribution to the
vector case from these diagrams, however, non-vanishing, finite
results are obtained for $c_a$, $c_s$ and $c_p$. 
For the scalar and the pseudo-scalar currents only the heavy quark is
running in the closed fermion loop. All other quark flavours are
suppressed by the light quark mass.
This is different for the axial-vector coupling. Here we consider the
effective current formed by the top and bottom quark field
\begin{eqnarray}
  j_a^\mu &=& \bar{t} \gamma^\mu \gamma_5 t -
  \bar{b} \gamma^\mu \gamma_5 b\,,
\end{eqnarray}
which ensures the cancellation of the anomaly-like contributions. For
the same reason the contributions from the remaining light quarks
cancel.

In the analytical results given below the contributions from the
singlet diagrams are marked
separately. At this point we only want to mention that in the
axial-vector 
and pseudo-scalar case $\gamma_5$ was treated according to the
prescription of Ref.~\cite{Larin:1993tq}. In practice this means that
we perform the replacements
\begin{eqnarray}
  \gamma^\mu\gamma_5 &\to& 
  \frac{i}{3!}\epsilon^{\mu\nu\rho\sigma} \gamma_\nu\gamma_\rho\gamma_\sigma
  \,,\nonumber\\
  \gamma_5           &\to& 
  \frac{i}{4!}\epsilon^{\mu\nu\rho\sigma} 
  \gamma_\mu\gamma_\nu\gamma_\rho\gamma_\sigma
  \label{eq::g5}
  \,,
\end{eqnarray}
strip off the $\epsilon$ tensor and deal with the objects with three
and four indices, respectively. 
The corresponding projectors are obtained by performing the
replacements of Eq.~(\ref{eq::g5}) in Eqs.~(\ref{eq::proj1})
and~(\ref{eq::proj2}) which makes them more complicated.
However, the very calculation is in close analogy to the non-singlet
case. After summing all two-loop contributions one obtains a finite
result.

Note that for the non-singlet contributions it is save to use
anti-commuting $\gamma_5$. Actually, the treatment according to 
Ref.~\cite{Larin:1993tq} leads to a wrong result. This is due to the 
infra-red divergences which are absent in the singlet diagrams.

An alternative method to perform the calculation of the vertex
corrections
is based on the evaluation of the tensor integrals. Here, we used the
T-operator method of Ref.~\cite{Tarasov:1996br} to reduce the tensor
integrals to products of the metric tensor and
external momenta and scalar integrals 
with shifted space-time dimension. The resulting Dirac structures were
further simplified and rewritten in terms of NRQCD fields.

In addition to the bare two-loop diagrams we have to take into account
the one-loop renormalization contribution from the heavy quark mass,
which we renormalize on-shell, and the strong coupling renormalized in
the $\overline{\rm MS}$ scheme.

We want to mention that all contributions have been evaluated 
for general gauge parameter $\xi$. The final results 
for the matching coefficients are all independent of $\xi$ which 
constitutes an important check on our calculation.

Let us in the following present our results and compare with the
literature.
The two-loop matching coefficient for the vector current 
has been computed almost ten years 
ago~\cite{Czarnecki:1997vz,Beneke:1997jm}. We confirmed these results
and provide for completeness the analytical expressions
\begin{eqnarray}
  c_v &=& 1 - 2 \frac{\alpha_s(m)}{\pi} C_F +
  \left(\frac{\alpha_s(m)}{\pi}\right)^2 \left[ C_F T \left(
    \frac{11}{18} n_l + \frac{22}{9} - \frac{4}{3} \zeta_2
    \right) \right. \nonumber\\
    && + C_F^2 \left(
    \frac{23}{8} - \frac{79}{6} \zeta_2 + 6 \zeta_2 \ln 2 -
    \frac{1}{2} \zeta_3 - \zeta_2 \ln\frac{\mu^2}{m^2}
    \right) \nonumber\\
    && \left. + C_F C_A \left( -\frac{151}{72} + \frac{89}{24} \zeta_2 -
    5 \zeta_2 \ln 2 - \frac{13}{4} \zeta_3 - \frac{3}{2}
    \zeta_2 \ln\frac{\mu^2}{m^2} \right) \right]
  \,,
\end{eqnarray}
where $C_A = N_c$ and $C_F = (N_c^2 - 1)/(2 N_c)$ are the Casimir
operators of the adjoint and fundamental representation of SU($N_c$),
respectively, $T = 1/2$, and $n_l$ is the number of massless
quarks. $\zeta_n$ denotes Riemann's zeta-function.
The one-loop result can already be found in Ref.~\cite{KalSar}.
The anomalous dimension of the effective vector current, which is
related to $\tilde{Z}_v$ through 
$\gamma_v = \frac{{\rm d} \ln \tilde{Z}_v}{{\rm d} \ln \mu}$, 
reads
\begin{eqnarray}
  \gamma_v &=& -\left(\frac{\alpha_s}{\pi}\right)^2\left( 2
  C_F^2 + 3 C_F C_A \right) \zeta_2
  \,.
\end{eqnarray}

Our results for the two-loop matching coefficients $c_a$, $c_s$ and
$c_p$ are given by
\begin{eqnarray}
  c_a &=& 1 - \frac{\alpha_s(m)}{\pi} C_F +
  \left(\frac{\alpha_s(m)}{\pi}\right)^2 \left[ C_F T \left(
    \frac{7}{18} n_l + \frac{20}{9} - \frac{4}{3} \zeta_2
    \right) \right. \nonumber\\
    && + C_F^2 \left(
    \frac{23}{24} - \frac{27}{4} \zeta_2 + \frac{19}{4} \zeta_2 \ln 2 -
    \frac{27}{16} \zeta_3 - \frac{5}{4} \zeta_2 \ln\frac{\mu^2}{m^2}
    \right) \nonumber\\
    && \left. + C_F C_A \left( -\frac{101}{72} + \frac{35}{24} \zeta_2 -
    \frac{7}{2} \zeta_2 \ln 2 - \frac{9}{8} \zeta_3 - \frac{1}{2}
    \zeta_2 \ln\frac{\mu^2}{m^2} \right)
    + C_F T X^{(a)}_{\rm sing}
    \right]
  \,,\nonumber\\
  c_s &=& 1 - \frac{1}{2} \frac{\alpha_s(m)}{\pi} C_F +
  \left(\frac{\alpha_s(m)}{\pi}\right)^2 \left[ C_F T \left(
    -\frac{5}{36} n_l + \frac{121}{36} - 2 \zeta_2
    \right) \right. \nonumber\\
    && + C_F^2 \left(
    \frac{5}{16} - \frac{37}{8} \zeta_2 + 3 \zeta_2 \ln 2 -
    \frac{11}{4} \zeta_3 - 2 \zeta_2 \ln\frac{\mu^2}{m^2}
    \right) \nonumber\\
    && \left. + C_F C_A \left( \frac{49}{144} + \frac{1}{8} \zeta_2 -
    3 \zeta_2 \ln 2 - \frac{5}{4} \zeta_3 - \frac{1}{2}
    \zeta_2 \ln\frac{\mu^2}{m^2} \right)
    + C_F T X^{(s)}_{\rm sing}
    \right]
  \,,\nonumber\\
  c_p &=& 1 - \frac{3}{2} \frac{\alpha_s(m)}{\pi} C_F +
  \left(\frac{\alpha_s(m)}{\pi}\right)^2 \left[ C_F T \left(
    \frac{1}{12} n_l + \frac{43}{12} - 2 \zeta_2
    \right) \right. \nonumber\\
    && + C_F^2 \left(
    \frac{29}{16} - \frac{79}{8} \zeta_2 + 6 \zeta_2 \ln 2 - \frac{9}{2}
    \zeta_3 - 3 \zeta_2 \ln\frac{\mu^2}{m^2} \right) \nonumber\\
    && \left. + C_F C_A \left( -\frac{17}{48} + \frac{17}{8} \zeta_2 - 6
    \zeta_2 \ln 2 - 3 \zeta_3 - \frac{3}{2} \zeta_2 \ln\frac{\mu^2}{m^2}
    \right)
    + C_F T X^{(p)}_{\rm sing}
    \right]
  \,.
  \label{eq::c_asp}
\end{eqnarray}
The one-loop result for $c_p$ can already be found in
Ref.~\cite{Braaten:1995ej};
the two-loop coefficients of Eq.~(\ref{eq::c_asp}) are new. 
They constitute our main result.
The one-loop coefficients can be easily obtained from the one-loop
on-shell vertex corrections with arbitrary momentum squared of the
external current, $s$. In the analytic expressions
it is straightforward to perform the limit where the velocity of the 
produced quarks is small. After subtracting the leading term, which
corresponds to the Coulomb singularity, one remains with the result
for the matching coefficients~\cite{Chetyrkin:1997mb}.
At two loops this simple trick does not work any more and the
calculation has to be performed from scratch as has been done in this letter.

The contributions from the singlet diagrams correspond to
\begin{eqnarray}
  X^{(a)}_{\rm sing} &=& -\frac{23}{12} \zeta_2 + 4 \zeta_2 \ln 2 - 2
  \ln 2 + \frac{2}{3} \ln^2 2 + i\pi \left( 1 - \frac{2}{3} \ln 2 \right)
  \,,\nonumber\\
  X^{(s)}_{\rm sing} &=& \frac{2}{3} - \frac{29}{12} \zeta_2 +
  4 \zeta_2 \ln 2 - \ln 2 
  + i\frac{\pi}{2}
  \,,\nonumber\\
  X^{(p)}_{\rm sing} &=& \frac{5}{4} \zeta_2 + 3 \zeta_2 \ln 2 -
  \frac{21}{8} \zeta_3 + i\pi \frac{3}{4} \zeta_2 
  \,.
  \label{eq::res_sing}
\end{eqnarray}
$X^{(s)}_{\rm sing}$ and $X^{(p)}_{\rm sing}$ receive only contributions
from diagrams which are finite and contain only the heavy quark.
The corresponding result holds both for top and bottom quarks.
This is different in the case of $X^{(a)}_{\rm sing}$. Actually, the
result in Eq.~(\ref{eq::res_sing}) corresponds to
the case where top quarks are considered in the final state.
Note that $X^{(a)}_{\rm sing}$ receives contributions from diagrams
with top and bottom quarks in the closed triangle loop
(cf. Fig.~\ref{fig::diags}(e)). Taken separately they are divergent,
however, the sum is finite. If one considers bottom quarks in the
final state one still has to consider top and bottom quarks in the
closed triangle loop. Again only the sum of all diagrams is finite
with the result
\begin{equation}
  X^{(a)}_{\rm sing} = \frac{55}{24} + \frac{19}{12} \zeta_2 - 4 \zeta_2
  \ln 2 - \frac{3}{4} \ln \frac{m_b^2}{m_t^2}
  + {\cal O}\left(\frac{m_b^2}{m_t^2}\right)
  \,.
  \label{eq::res_sing2}
\end{equation}
The diagram in Fig.~\ref{fig::diags}(e) for axial-vector coupling and
with bottom in the final state was also considered in
Ref.~\cite{Kniehl:1989bb,Kniehl:1989qu}, 
for abritrary values of $s$ and $m_t$, but for $m_b=0$ so that no
direct comparison with Eq.~(\ref{eq::res_sing2}) is possible.

As mentioned above, it is possible to extract the result for $c_p$
from the zero component of the axial-vector current. This is quite
evident in the non-singlet case. However, for the singlet contribution
this check is highly non-trivial since in this approach $c_p$ is
obtained from diagrams both with top and bottom quarks in the closed
triangle whereas in the direct calculation only one type of quarks
appears.

The singlet results of
Eqs.~(\ref{eq::res_sing}) and~(\ref{eq::res_sing2}) 
and the fermionic contributions of Eq.~(\ref{eq::c_asp})
are in agreement with
Ref.~\cite{Bernreuther:2004th,Bernreuther:2005rw,Bernreuther:2005gw}
where the off-shell
contributions have been considered. Since they do not develop
an infrared singularity the limit $s\to 4m^2$ can be performed. This
is different in the case of the non-fermionic contributions where due to
the infrared divergence the off-shell
results~\cite{Bernreuther:2004ih,Bernreuther:2004th,Bernreuther:2005gw} 
cannot be used in order to obtain the matching coefficients.

For completeness we also provide the result for the anomalous
dimensions corresponding to Eq.~(\ref{eq::c_asp}) which read
\begin{eqnarray}
  \gamma_a &=& -\left(\frac{\alpha_s}{\pi}\right)^2\left(
  \frac{5}{2} C_F^2 + C_F C_A \right) \zeta_2
  \,,\nonumber\\
  \gamma_s &=& -\left(\frac{\alpha_s}{\pi}\right)^2\left( 4
  C_F^2 + C_F C_A \right) \zeta_2
  \,,\nonumber\\
  \gamma_p &=& -\left(\frac{\alpha_s}{\pi}\right)^2\left( 6
  C_F^2 + 3 C_F C_A \right) \zeta_2
  \,.
  \label{eq::gamma_asp}
\end{eqnarray}
The result for $\gamma_p$ agrees with the one extracted from
Ref.~\cite{Hoang:2005dk}.

We want to mention that the coefficient $c_p$ has been considered in
Ref.~\cite{Onishchenko:2003ui} in the context of the $B_c$ meson.
The latter consists of two heavy quarks, however, with different
masses. This makes the calculation significantly more difficult since
two instead of one mass scale appear in the integrals. In
Ref.~\cite{Onishchenko:2003ui} the reduction to master integrals has
been performed exactly whereas the latter have been evaluated in the
limit $m_c\ll m_b$
so that a comparison with the present analysis is not
possible.

To summarize, in this paper we computed the two-loop matching
coefficients between QCD and NRQCD for an axial-vector, scalar and 
pseudo-scalar current. Furthermore, we performed an independent check
of the matching coefficient in the vector case.
The latter contributes to the second order result of the 
threshold production of top quark pairs. The result for the axial-vector 
current only contributes to the fourth-order analysis which is
currently still out of reach.

\bigskip
\noindent
{\bf Acknowledgments.}\\
We would like to thank K.G. Chetyrkin, A.A. Penin, and V.A. Smirnov for
useful discussions. J.H.P. would like to thank S. Bekavac for
discussions about Mellin-Barnes integrals. This work was supported by
the ``Impuls- und Vernetzungsfonds'' of the Helmholtz Association,
contract number VH-NG-008 and the SFB/TR~9. The Feynman diagrams were
drawn with {\tt JaxoDraw} \cite{Binosi:2003yf}.

%- }}}

%- {{{ bibliography:

%- }}}

\end{document}